\documentclass[hyper]{JHEP} 

\usepackage{epsfig}

\def\be{\begin{equation}}
\def\ee{\end{equation}}
\def\ba{\begin{array}}
\def\ea{\end{array}}
\def\bea{\begin{eqnarray}}
\def\eea{\end{eqnarray}}
\def\nn{\nonumber\\}

\def\ct{\cite}
\def\la{\label}
\def\eq#1{Eq. (\ref{#1})}

\def\a{\alpha}
\def\b{\beta}

\def\G{\Gamma}

\def\D{\Delta}

\def\ph{\phi}

\def\ps{\psi}

\def\k{\kappa}
\def\l{\lambda}

\def\n{\nu}
\def\th{\theta}

\def\s{\sigma}

\def\o{\omega}

\newcommand\fverb{\setbox\pippobox=\hbox\bgroup\verb}

\newcommand\fverbdo{\egroup\medskip\noindent%

            \fbox{\unhbox\pippobox}\ }

\newcommand\fverbit{\egroup\item[\fbox{\unhbox\pippobox}]}

\newbox\pippobox

\title{Magnon like solutions for strings in I-brane background}

\author{J. Kluso\v{n}$^a$, Bum-Hoon Lee $^b$, Kamal L. Panigrahi $^c$ 
and Chanyong Park $^b$\\
$^a$ Department of Theoretical Physics and Astrophysics\\
~~Faculty of Science, Masaryk University\\
~~Kotl\'{a}\v{r}sk\'{a} 2, 611 37, Brno, Czech Republic\\

$^b$ Center for Quantum Spacetime (CQUeST), Sogang University, \\
~~Seoul 121-742, Korea \\

$^c$ Department of Physics, Indian Institute of Technology Guwahati,\\
~~Guwahati-781 039, India \\

E-mail: \email{klu@physics.muni.cz, bhl@sogang.ac.kr, panigrahi@iitg.ernet.in, cyong21@sogang.ac.kr}}

\preprint{{0806.3879 [hep-th]}}

 \abstract{We study the solutions for fundamental string rotating in 
a background generated by a 1+1 
dimensional intersection of two orthogonal stacks of fivebranes in type IIB 
string theory. We show the
 existence of magnon like solutions for the string moving simultaneously
in the two spheres
 in this background and find the relevant dispersion relation among the 
various conserved charges.}


\def\bz{\mathbf{z}}
\def\by{\mathbf{y}}

\begin{document}

\section{Introduction and Summary}\label{first}

AdS/CFT correspondence \ct{mal1} relates the spectrum of free
strings on AdS$_5\times$ S$^5$ with that of the spectrum of operator
dimensions in the ${\cal N}=4$ super Yang-Mills (SYM) in four
dimensions. This mapping is highly nontrivial and challenging. A
better understanding of this mapping will be to look at both the
gauge and gravity theories at certain limits such as large angular
momentum limit and then compare the spectrum. In the understanding
the gauge/gravity duality, an interesting observation is that the
${\cal N}=4$ SYM theory can be described by the integrable spin
chain model where the anomalous dimension of the gauge invariant
operators were found \ct{zm,intg1,intg2,intg3,intg4,intg5,intg6}. It
was further noticed that the string theory also has as integrable
structure in the semiclassical limit and the anomalous dimension in
the ${\cal N}=4$ SYM can be derived from the relation between
conserved charges of the worldsheet solitonic string solution on the
dual string theory on AdS$_5 \times$ S$^5$. In this connection,
Hofman and Maldacena (HM) \ct{hm} considered a special limit where
the problem of determining the spectrum of both sides becomes rather
simple. The spectrum consists of an elementary excitation known as
magnon which propagate with a conserved momentum $p$ along the long
spin chain. In the dual formulation, the most important ingredient
is semiclassical string solutions, which can be mapped to long trace
operator with large energy and large angular momenta. Once this
connection was established, there was a lot of work devoted to
understanding of this correspondence (see for
example\cite{Dorey:2006dq}-\cite{David:2008yk}).

To understand the AdS/CFT like correspondence in a more general
background, it is interesting to find out dispersion relation among
various conserved charges in case of classical rotating strings in
the gravity side and then to look for the corresponding operators in
the dual theory. For example, the magnon like dispersion relation in
NS5-brane background was considered in \cite{Kluson:2007qu}.

An interesting background configuration is the I-brane background
\cite{Itzhaki:2005tu}, which is the intersection of two stacks of
NS5-branes in type II string theory on $R^{1,1}$. When all the five
branes are coincident, the near horizon geometry of this
configuration is given by \cite{Itzhaki:2005tu}
\footnote{See \cite{Lin:2005nh}
for related background and for the gauge theory results.} 
\be R^{2,1} \times
R_{\ph} \times SU(2)_{k_1} \times SU(2)_{k_2}, \label{i-brane}\ee
where $R_{\ph}$ is one combination of two radial directions away
from two sets of NS5-branes, $k_1$ and $k_2$ are the number of
NS5-branes in each stack. The coordinates of $R^{2,1}$ are $x^0,
x^1$ and one more combination of two radial directions. The two
$SU(2)$s corresponding to two angular three spheres corresponding to
$(R^4)_{2345}$ and $(R^4)_{6789}$. The very fact that
(\ref{i-brane}) is an exact solution to string equations of motion
allows us to obtain information about the intersecting brane system
more, which is not accessible via gauge theory. As it is clear from
(\ref{i-brane}) the exhibits a higher symmetry than the full
intersecting brane configuration. In particular, the combination of
radial directions away from the intersection that enters $R^{2,1}$
appears symmetrically with the other spatial directions, and the
background has a higher Poincare symmetry, $ISO(2, 1)$, than the
expected $ISO(1, 1)$. The holographic mapping between field theory
living on the I-brane and the corresponding bulk theory was studied
in \cite{Itzhaki:2005tu} \footnote{For some relevant works, see
\cite{Hung:2006nn,Berg:2006ng,Hung:2006jh,Grisa:2006tm,Antonyan:2006pg,
Antonyan:2006qy,Antonyan:2006vw,Kluson:2006wa,
Kluson:2005qq,Kluson:2005eb}}.  On the other hand it is also well
known from the study of AdS/CFT correspondence that it is possible
to derive more information about the boundary  CFT theory from the
study of the semiclassical string and D1-brane configurations in the
bulk of $AdS_5$ \footnote{For review, see
\cite{Tseytlin:2003ii,Tseytlin:2004xa,Plefka:2005bk}.}.

Motivated by the recent developments in the AdS/CFT correspondence,
we will investigate the classical string dynamics in the I-brane
background to understand the rotating string solutions. These string
solutions correspond to multispin string solutions on the I-brane
background and solitonic string solution in the worldsheet point of
view. Due to the lack of sufficient knowledge about the theory on
the worldvolume of the I-brane we can't compare them with the dual
theory. However the knowledge about the semiclassical rotating
string on the I-brane certainly gives information about the possible
nature of operators in some dual theory.

In this background, the solitonic
string solution which are static and
uniformly wrapping the two transverse
spheres, and the dispersion relation
among various conserved charges were
studied in \ct{Kluson:2007st}. Here, we
will generalize this solution to the
rotating one in the transverse spheres
and find the dispersion relation among
various charges. To do so, we will
first solve the equations of motion for
a string rotating simultaneously on
both the spheres and then by using the
Virasoro constraints the dispersion
relation will be expressed in terms of
various conserved charges that the
background obeys in general. Usually,
in case of the giant magnon on $R
\times S^3$, the energy and one of the
angular momenta for one giant magnon
are infinite but their difference is
finite. In the I-brane background, on
the other hand, which has also $R
\times S^3$ or $R \times S^3 \times
S^3$, the string soliton is composed of
multiple magnon-like solutions and one
of them, we call this magnon-like
solution, has a similar shape to the
giant magnon on $R \times S^3$.
However, the dispersion relation for
magnon-like one is different in that it
contains additional linear momenta in
two radial directions and is described
by the finite conserved charges. We
would like to note that on the I-brane
background all conserved charges are
regularized, so have finite values. If
we choose the range of the world sheet
string coordinate as $-\infty < \s <
\infty$ like HM case \ct{hm} or $-\pi/2
< \s < \pi/2$ , the string soliton
becomes a combination of infinite
magnon-like solutions or finite
numbers. For the closed string case,
since $-\pi < \s < \pi$ the solitonic
string is given by the finite numbers
of the magnon-like solutions.

We would like to mention that in general getting magnon/spike
solutions in I-brane background is rather cumbersome. However, we
will present in this paper, a parameter space of solutions where
there exists a magnon like shape when we restrict the motion of the
string along both the spheres. It would be certainly interesting to
find out more general rotating string solutions in this backgrounds.
A greater challenge will be to find out the dual operators in the
worldvolume theory which correspond to the semiclassical string
solutions presented in this paper.

The rest of the paper is organized as follows. In section 2, we
present the background solution corresponding to the intersection of
two five-branes on $R^{1,1}$ in type IIB string theory and study the
equations of motion and Virasoro constraints of the fundamental
string rotating simultaneously along both the spheres. Section-3
devoted to the study the rotating string solution interpreted as a
single magnon like solution while the motion is restricted to only
one sphere. We present the corresponding dispersion relation along
various charges and interpret that solution as a single magnon
solution. Further we present a more general solution when the string
moves simultaneously along both the spheres. We analyze the results
in a particular parameter space of solutions. Finally in section-4,
we present our conclusions.

\section{F-string in the background
of I-brane}\label{third} In this section we will study the dynamics
of fundamental string in the  background studied in
\cite{Itzhaki:2005tu}. Namely, we consider  the intersection of two
stack of NS5-branes on $R^{1,1}$. We have $k_1$ number of NS5-branes
extended in $(0,1,2,3,4,5)$ directions and another set of $k_2$
number of NS5-branes extended in $(0,1,6,7,8,9)$ directions. Let us
define
\begin{eqnarray}
\by=(x^2,x^3,x^4,x^5) \ , \nonumber \\
\bz=(x^6,x^7,x^8,x^9) \ .
\end{eqnarray}
We have $k_1$ NS5-branes localized at the points $\bz_n \
n=1,\dots,k_1 , $ and $k_2$ NS5-branes localized at the points
$\by_a \ , a=1\dots,k_2$. Every pairs of fivebranes from different
sets intersect at different point $(\by_a,\bz_n)$. The supergravity
background corresponding to this configuration takes the form
\begin{eqnarray}\label{bg}
\Phi(\bz,\by) &=& \Phi_1(\bz)+
\Phi_2(\by) \ , \nonumber \\
g_{\mu\nu}&=&\eta_{\mu\nu} \ ,
\ (\mu,\nu=0,1) \ , \nonumber \\
g_{\alpha\beta}&=&e^{2(\Phi_2-
\Phi_2(\infty))}\delta_{\alpha\beta} , \nn
\mathcal{H}_{\alpha\beta\gamma}&=&
-\epsilon_{\alpha\beta\gamma\delta}
\partial^\delta \Phi_2 \ , \ (\alpha,\beta,\gamma,\delta=
2,3,4,5 ) \ , \nonumber \\
g_{pq}&=&e^{2(\Phi_1-\Phi_1(\infty))}
\delta_{pq} \ , \nn
\mathcal{H}_{pqr}&=&-\epsilon_{pqrs}
\partial^s\Phi_1 \ , \ \ ( p,q,r,s=6,7,8,9 )\ ,
\end{eqnarray}
where $\Phi$ on the
first line means the dilaton
and where
\begin{eqnarray}
e^{2(\Phi_1-
\Phi_1(\infty))}=1+
\sum_{n=1}^{k_1}
\frac{l_s^2}{|\bz-\bz_n|^2} \ ,
\nonumber \\
e^{2(\Phi_2-
\Phi_2(\infty))}=1+\sum_{a=1}^{k_2}
\frac{l_s^2}{|\by-\by_a|^2} \ .
\end{eqnarray}
Our goal is to find solutions for rotating string in this background
when $\bz_n=\by_a=0$. To simplify our notation let us denote
\begin{equation}
e^{2(\Phi_1-\Phi_1(\infty))}=H_1(\bz)\ , \ \ \
e^{2(\Phi_2-\Phi_2(\infty))}=H_2(\by)  \ ,
\end{equation}
where for coincident branes we have
\begin{equation}
H_1=1+\frac{k_1l_s^2} {|\bz|^2} \ , \ \ \
H_2=1+\frac{k_2l_s^2} {|\by|^2} \ .
\end{equation}
Let us now consider the probe brane in the near horizon limit where
\begin{equation}
\frac{k_1l_s^2}
{|\bz|^2}\gg 1 \ , \ \ \
\frac{k_2l_s^2}
{|\by|^2}\gg 1 \
\end{equation}
so that  we can write
\begin{equation}
H_1=\frac{\lambda_1} {r^2_1} \ , \quad
\lambda_1=k_1l_s^2 \ , \quad
H_2=\frac{\lambda_2} {r^2_2} \ , \quad
\lambda_2=k_2l_s^2 \ .
\end{equation}
Then the metric takes the form
\begin{eqnarray}\label{NS5bac}
ds^2=-dt^2+\frac{\lambda_1}{r_1^2}dr_1^2+
\frac{\lambda_2}{r_2^2}dr_2^2+
\lambda_1d\Omega^{(3)}_1+
\lambda_2d\Omega^{(3)}_2 \ ,
\end{eqnarray}
where $d\Omega_1^{(3)}$ and $d\Omega_2^{(3)}$ correspond to the line
elements on the unit sphere. To describe them better we introduce
the following coordinates
\begin{eqnarray}
x^2+ix^3&=&r_1 \cos
\theta_1e^{i\phi_1}\ ,
\quad
x^4+ix^5=r_1\cos \theta_1
e^{i\psi_1}
\ ,
\nonumber \\
x^6+ix^7&=&r_2 \cos
\theta_2e^{i\phi_2}
 \ , \quad
x^8+ix^9=r_2\cos \theta_2
e^{i\psi_2} \nonumber \\
\end{eqnarray}

so that
\begin{eqnarray}    \la{NS5bacd}
d\Omega^{(3)}_1&=& d\theta^2_1+
\sin^2\theta_1 d \phi_1^2+
 \cos^2\theta_1 d\psi_1^2 , \nonumber \\
   b_{\phi_1\psi_1}&=&
   \lambda_1\cos^2\theta_1 \ ,
\quad   0 < \theta_1 < \frac{\pi}{2} \
, \quad 0 = \phi_1,\psi_1 <
  2\pi \ ,
  \nonumber \\
d\Omega^{(3)}_2&=& d\theta^2_2+
\sin^2\theta_2 d \phi_2^2+
 \cos^2\theta_2 d\psi_2^2 , \nonumber \\
    b_{\phi_2\psi_2}&=&
  \lambda_2\cos^2\theta_2 \ ,
 \quad 0 < \theta_2 < \frac{\pi}{2} \ ,
 \quad
  0 = \phi_2,\psi_2 < 2\pi \ .
\end{eqnarray}
As usual our starting point is the Polyakov form of the string
action in the background (\ref{NS5bac})
\begin{eqnarray}\label{actPol}
S&=&-\frac{1}{4\pi \alpha'}
\int_{-\pi/2}^{\pi/2} d\sigma d\tau
[\sqrt{-\gamma}\gamma^{\alpha\beta}
g_{MN}\partial_\alpha x^M\partial_\beta x^N
-e^{\alpha\beta}
\partial_\alpha x^M\partial_\beta x^N b_{MN}]+
\nonumber \\
&+&\frac{1}{4\pi}\int_{-\pi/2}^{\pi/2}
d\sigma d\tau \sqrt{-\gamma}
R\Phi \ ,
\end{eqnarray}
 where $\gamma^{\alpha\beta}$ is a
world-sheet metric and $R$ is its Ricci scalar. Further,
$e^{\alpha\beta}$ is defined as $e^{01}=-e^{10}=1$. Finally, the
modes $x^M, M=0,\dots,9$ parameterize the embedding of the string in
the background (\ref{NS5bac}). The variation of the action
(\ref{actPol}) with respect to $x^M$ implies following equations of
motion
\begin{eqnarray}
-\frac{1}
{4\pi\alpha'}\sqrt{-\gamma}\gamma^{\alpha\beta}
\partial_K
g_{MN}\partial_\alpha x^M\partial_\beta x^N
+\frac{1}{2\pi\alpha'}\partial_\alpha[
\sqrt{-\gamma}\gamma^{\alpha\beta}
g_{KM}\partial_\beta x^M]-
\nonumber \\
-\frac{1}{2\pi\alpha'}
\partial_\alpha[
\epsilon^{\alpha\beta}
\partial_\beta x^M b_{KM}]
+\frac{1}{4\pi\alpha'}
\epsilon^{\alpha\beta}
\partial_\alpha x^M\partial_\beta x^N
\partial_K b_{MN}
+\frac{1}{4\pi}\partial_K \Phi
\sqrt{-\gamma}R=0 \ . \nonumber \\
\end{eqnarray}
Finally  the variation of the action
with respect to the metric implies the
constraints
\begin{eqnarray}\label{gravcons}
-\frac{4\pi}{\sqrt{-\gamma}}
\frac{\delta S}{\delta \gamma^{\alpha
\beta}}&=&
\frac{1}{\alpha'}
g_{MN}\partial_\alpha x^M\partial_\beta x^N-R_{\alpha
\beta}+
\nonumber \\
&+&(\nabla_\alpha \nabla_\beta x^M)
\partial_M \Phi+(\partial_\alpha x^M\partial_\beta
x^N)\partial_M\partial_N\Phi
-\nonumber \\
&-&\frac{1}{2}
\gamma_{\alpha\beta}
\left(\frac{1}{\alpha'}
\gamma^{\gamma\delta}
\partial_\gamma x^M\partial_\delta
x^N g_{MN}-R\Phi+2 \nabla^\alpha
\nabla_\alpha \Phi\right) \ . \nonumber
\\
\end{eqnarray}
As the first step let us introduce two
modes $\rho_1$ and $\rho_2$ defined
through the relations
\begin{equation}\label{RSp}
r_1=e^{\frac{\rho_1}{\sqrt{\lambda_1}}
}\ , \quad
r_2=e^{\frac{\rho_2}{\sqrt{\lambda_2}}}
\ .
\end{equation}
Then, following  \cite{Itzhaki:2005tu} we introduce two modes $ r,
y$ through the relation
\begin{equation}\label{phix2}
Qr=\frac{1}{\sqrt{\lambda_1}}\rho_1+
\frac{1}{\sqrt{\lambda_2}}\rho_2 \ ,
\quad
Qy=\frac{1}{\sqrt{\lambda_2}}\rho_1-
\frac{1}{\sqrt{\lambda_1}}\rho_2 \ ,
\end{equation}
where
\begin{equation}
Q=\frac{1}{\sqrt{\lambda}} \ ,
\frac{1}{\lambda}=\frac{1}{\lambda_1}+
\frac{1}{\lambda_2} \ .
\end{equation}
Note that the inverse transformations of (\ref{phix2}) take the
forms
\begin{eqnarray}\label{phili}
\rho_1=\frac{1}{\sqrt{\lambda_1+\lambda_2}}
\left(\sqrt{\lambda_1}y+\sqrt{\lambda_2}r
\right) \ , \nonumber \\
\rho_2=\frac{1}{\sqrt{\lambda_1+\lambda_2}}
\left(\sqrt{\lambda_1}r-
\sqrt{\lambda_2}y\right) \ .
\end{eqnarray}
Note that this result implies that the dilaton is a function of $r$
only
\begin{eqnarray}
\Phi&=&\Phi_1+\Phi_2=
\frac{1}{2}(H_1+H_2)+\Phi_1(\infty)+
\Phi_2(\infty)=\nonumber \\
&=& -\frac{1}{\sqrt{\lambda_1}}\rho_1
-\frac{1}{\sqrt{\lambda_2}}\rho_2+\Phi_0=-Qr+\Phi_0
\ .
\end{eqnarray}
With the help of the variables $r,y$ the action
for string in $I$-brane background takes the form
\begin{eqnarray}\label{actPol2}
S&=&-\frac{1}{4\pi \alpha'}
\int_{-\pi/2}^{\pi/2} d\sigma d\tau
[\sqrt{-\gamma}\gamma^{\alpha\beta}
(-\partial_\alpha t\partial_\beta t
+\partial_\alpha r\partial_\beta r+
\partial_\alpha y\partial_\beta y+\nonumber \\
&+&g_{mn}\partial_\alpha
x^m\partial_\beta x^n -e^{\alpha\beta}
\partial_\alpha x^m\partial_\beta x^n b_{mn}]-
\frac{1}{4\pi}\int_{-\pi/2}^{\pi/2} d\sigma
d\tau \sqrt{-\gamma} RQr \ ,
\end{eqnarray}
where $x^{m,n}$ label angular coordinates
corresponding to $S^3_1,S^3_2$ respectively.

Looking at the form of the background (\ref{NS5bac}) and
(\ref{NS5bacd}) we observe that the action (\ref{actPol2}) is
invariant under following transformations of fields
\begin{eqnarray}
t'(\tau,\sigma)&=&t(\sigma,\tau)+\epsilon_t \ ,
\nonumber \\
y'(\tau,\sigma)&=&y(\tau,\sigma)+\epsilon_y  \ ,
\nonumber \\
\psi'_{1}(\tau,\sigma)&=&
\psi_{1}(\tau,\sigma)+\epsilon_{\psi_1} \ ,
\nonumber \\
\psi'_{2}(\tau,\sigma)&=&
\psi_{2}(\tau,\sigma)+\epsilon_{\psi_2} \ ,
\nonumber \\
\phi'_{1}(\tau,\sigma)&=&
\phi_{1}(\tau,\sigma)+\epsilon_{\phi_1} \ ,
\nonumber \\
\phi'_{2}(\tau,\sigma)&=&
\phi_{2}(\tau,\sigma)+\epsilon_{\phi_2} \ ,
\end{eqnarray}
where $\epsilon_t,\epsilon_y,\epsilon_{\phi_1},\epsilon_{\phi_2},
\epsilon_{\psi_1},\epsilon_{\psi_2}$ are constants. Then it is
straightforward to determine corresponding conserved charges
\begin{eqnarray}\label{CGcon}
P_t&=&-\frac{1}{2\pi\alpha'}
\int_{-\pi/2}^{\pi/2} d\sigma
\sqrt{-\gamma}\gamma^{\tau\alpha}
\partial_\alpha t \ , \nonumber \\
P_{\psi_1}&=&
\frac{1}{2\pi\alpha'}
\int_{-\pi/2}^{\pi/2} d\sigma
[\sqrt{-\gamma}\gamma^{\tau\alpha}
g_{\psi_1\psi_1}\partial_\alpha \psi_1
-\partial_\sigma \phi_1 b_{\phi_1\psi_1}] \ ,
\nonumber \\
P_{\psi_2}&=&
\frac{1}{2\pi\alpha'}
\int_{-\pi/2}^{\pi/2} d\sigma
[\sqrt{-\gamma}\gamma^{\tau\alpha}
g_{\psi_2\psi_2}\partial_\alpha \psi_2
-\partial_\sigma \phi_2 b_{\phi_2\psi_2}] \ ,
\nonumber \\
P_{\phi_1}&=&
\frac{1}{2\pi\alpha'}
\int_{-\pi/2}^{\pi/2} d\sigma
[\sqrt{-\gamma}\gamma^{\tau\alpha}
g_{\psi_1\psi_1}\partial_\alpha \psi_1
+\partial_\sigma \psi_1 b_{\phi_1\psi_1}] \ ,
\nonumber \\
P_{\phi_2}&=&
\frac{1}{2\pi\alpha'}
\int_{-\pi/2}^{\pi/2} d\sigma
[\sqrt{-\gamma}\gamma^{\tau\alpha}
g_{\psi_2\psi_2}\partial_\alpha \psi_2
+\partial_\sigma \psi_2 b_{\phi_2\psi_2}] \ ,
\nonumber \\
P_{y}&=&
\frac{1}{2\pi\alpha'}
\int_{-\pi/2}^{\pi/2} d\sigma
[\sqrt{-\gamma}\gamma^{\tau\alpha}
g_{yy}\partial_\alpha y] \ .
\end{eqnarray}
Note that $P_t$ is related to the energy as $P_t=-E$. In
\cite{Kluson:2007st} the homogeneous string and D1-brane solutions
in I-brane background have been studied \footnote{Dynamics of
D1-brane probe in given background was studied in
\cite{Kluson:2005eb,Kluson:2005qq}.}. It was argued that it is
necessary to find the configuration when string moves on both two
spheres simultaneously. For that reason we have to consider an
ansatz where string moves simultaneously on both the spheres
$S^{3}_1$ and $S^{3}_2$ as follows
\begin{eqnarray}
t&=&t(\tau), \quad r= r(\tau), \quad
y=y(\tau), \quad
\nonumber \\
\theta_1&=&\theta_1(m) \ , \quad
\psi_1=\omega_1\tau+g_1(m) \ , \quad
\phi_1=\nu_1\tau + h_1(m) \ ,
\nonumber \\
\theta_2&=&\theta_2(m) \ , \quad
\psi_2=\omega_2\tau+g_2(m) \ , \quad
\phi_2=\nu_2\tau + h_2(m) \ , \nonumber
\\
\end{eqnarray}
where
\begin{equation}\label{defm}
m=\alpha \sigma+\beta \tau
\end{equation}
and we also consider solution in the
conformal gauge $\gamma^{\alpha\beta}=
\eta^{\alpha\beta}$. In this gauge the
constraints (\ref{gravcons}) that now
follow from the variation of the action
(\ref{actPol2}) take simpler forms
\begin{eqnarray}\label{Tcon}
T_{\sigma\sigma}&=&-4\pi \frac{\delta
S}{\delta \gamma^{\sigma \sigma}}=
\frac{1}{2\alpha'}
(g_{MN}\partial_\sigma
x^M\partial_\sigma x^N
+g_{MN}\partial_\tau x^M\partial_\tau
x^N)- Q\partial_\tau^2 \rho \ ,
\nonumber \\
T_{\tau\tau}&=&-4\pi \frac{\delta
S}{\delta \gamma^{\tau \tau}}=
\frac{1}{2\alpha'}
(g_{MN}\partial_\sigma
x^M\partial_\sigma x^N
+g_{MN}\partial_\tau x^M\partial_\tau
x^N)- Q\partial_\sigma^2 \rho \ ,
\nonumber \\
T_{\tau\sigma}&=&-4\pi \frac{\delta
S}{\delta \gamma^{\sigma\tau }}=
\frac{1}{\alpha'} g_{MN}\partial_\sigma
x^M\partial_\tau x^N
-Q\partial_\sigma\partial_\tau \rho \ .
\end{eqnarray}
Further, in conformal gauge the
equations of motion for $t,y,r$ take
the form
\begin{eqnarray}
\partial_\alpha[\eta^{\alpha\beta}\partial_\beta
t]=0 \ , \quad
\partial_\alpha[\eta^{\alpha\beta}\partial_\beta
r]=0 \ , \quad
\partial_\alpha[\eta^{\alpha\beta}\partial_\beta
y]=0 \
\end{eqnarray}
that have solutions
\begin{equation}
t=\kappa \tau \ , \quad r=v_r \tau+r_0
\ , \quad  y=v_y\tau+y_0 \
\end{equation}
for constants $\kappa,v_y,v_r,r_0,y_0$.

Now we are going to study the dynamics
of fundamental strings on both three
$S^{3}_1,S^{3}_2$.
We start then with the equations of
motion for $\phi_1$ and $\psi_1$.
It can be easily shown that these
equations imply two differential
equations
\begin{eqnarray}\label{hg1}
h_1'&=&\frac{1}
{\lambda_1\sin^2\theta_1(\alpha^2-\beta^2)}
(C_1-\lambda_1\alpha\omega_1\cos^2\theta_1+\lambda_1\beta
\nu_1 \sin^2\theta_1) \ , \nonumber \\
 g'_1&=&\frac{1}{
(\alpha^2-\beta^2)\lambda_1\cos^2\theta_1}
(D_1+\lambda_1\beta\omega_1\cos^2\theta_1+
\lambda_1\nu_1\alpha \cos^2\theta_1) \
.
\nonumber \\
\end{eqnarray}
where $h'(m)=\frac{dh}{dm}$ and where
$C_1,D_1$ are integration constants.
In the same way we consider the
equation of motion for $\psi_2,\phi_2$
with the result
\begin{eqnarray}\label{hg2}
h'_2&=&\frac{1}{(\alpha^2-\beta^2)
\lambda_2\sin^2\theta_1} (C_2+\lambda_2
\sin^2\theta_2 \beta \nu_2
-\omega_2\alpha
\lambda_2\cos^2\theta_2) \ ,  \nonumber \\
 g'_2&=&\frac{1}{
(\alpha^2-\beta^2)\lambda_2\cos^2\theta_2}
(D_2+\lambda_2\beta\omega_2\cos^2\theta_2+
\lambda_2\nu_2\alpha \cos^2\theta_2) \
, \nonumber \\
\end{eqnarray}
where again $C_2,D_2$ are integration
constants.
Let us now start to solve the Virasoro constraints for the given
model. The constraint $T_{\tau\sigma}=0$ implies
\begin{eqnarray}
& & g_{\phi_1\phi_1}\alpha
h'_1(\nu_1+\beta h'_1)+
g_{\psi_1\psi_1}\alpha
g'_1(\omega_1+\beta g'_1)
+g_{\theta_1\theta_1}\alpha \beta
\theta'^2_1 +\nonumber \\
&& + g_{\phi_2\phi_2}\alpha
h'_2(\nu_2+\beta h'_2)+
g_{\psi_2\psi_2}\alpha
g'_2(\omega_2+\beta g'_2) +g_{\theta_2
\theta_2}\alpha\beta\theta'^2_2=0 \ .
\end{eqnarray}
On the other hand the Virasoro
constraints
$T_{\tau\tau}=T_{\sigma\sigma}=0$ take
the form
\begin{eqnarray}
0 &=& g_{\phi_1\phi_1}[(\nu_1+\beta
h'_1)^2+\alpha^2h'^2_1]+
g_{\psi_1\psi_1}[(\omega_1+\beta
g'_1)^2+\alpha^2 g'^2_1]+
g_{\theta_1\theta_1}(\alpha^2+\beta^2)\theta'^2_1+
\nonumber \\
&& g_{\phi_2\phi_2}[(\nu_2+\beta h'_2)^2+
\alpha^2 h'^2_2]+ g_{\psi_2\psi_2}
[(\omega_2+\beta g'_2)^2+ \alpha^2
g'^2_2]+g_{\phi_2\phi_2}
(\alpha^2+\beta^2)\theta'^2_2
-\nonumber \\
&& -\kappa^2+v_r^2+v_y^2 \ .  \nonumber \\
\end{eqnarray}
Now, if we combine these constraints as
$-\frac{(\alpha^2+\beta^2)}{\alpha\beta}T_{\sigma\tau}
+T_{\tau\tau}$ we obtain
\begin{equation}\label{const}
0= -\nu_1 C_1-\omega_1 D_1
-\nu_2C_2-\omega_2 D_2
+\beta(-\kappa^2+v_r^2+v_y^2)  \ .
\end{equation}
Then  if we use (\ref{hg1}), (\ref{hg2})
together with (\ref{const}) in the
constraint $T_{\tau\tau}=0$
 we obtain
\begin{eqnarray}    \label{mas1}
&& (\beta^2+\alpha^2)(\lambda_1\theta'^2_1+
\lambda_2\theta'^2_2) \nn
&& =
\frac{(\alpha^2+\beta^2)^2}
{(\alpha^2-\beta^2)^2}(
\kappa^2-v_r^2-v_y^2)-\frac{(\alpha^2+\beta^2)\alpha^2}{
(\alpha^2-\beta^2)^2}(g_{\phi_1\phi_1}\nu_1^2+
g_{\psi_1\psi_1}\omega_1^2)
(\frac{g_{\phi_1\phi_1}g_{\psi_1\psi_1}+
b_{\phi_1\psi_1}^2}{g_{\phi_1\phi_1}
g_{\psi_1\psi_1}})\nonumber \\
&&-\frac{\alpha^2+\beta^2}
{(\alpha^2-\beta^2)^2}[\frac{C_1^2}{g_{\phi_1\phi_1}}+
\frac{D_1^2}{g_{\psi_1\psi_1}}+
 +2\alpha
\frac{b_{\phi_1\psi_1}}
{g_{\psi_1\psi_1}g_{\phi_1\phi_1}}
(D_1\nu_1 g_{\phi_1\phi_1}- C_1
\omega_1 g_{\psi_1\psi_1})]
\nonumber \\
&&- \frac{(\alpha^2+\beta^2)\alpha^2}{
(\alpha^2-\beta^2)^2}(g_{\phi_2\phi_2}\nu_2^2+
g_{\psi_2\psi_2}\omega_2^2)
(\frac{g_{\phi_2\phi_2}g_{\psi_2\psi_2}+
b_{\phi_2\psi_2}^2}{g_{\phi_2\phi_2}
g_{\psi_2\psi_2}})-\nonumber \\
&&-\frac{\alpha^2+\beta^2}
{(\alpha^2-\beta^2)^2}[\frac{C_2^2}{g_{\phi_2\phi_2}}+
\frac{D_2^2}{g_{\psi_2\psi_2}}
 +2\alpha
\frac{b_{\phi_2\psi_2}}
{g_{\psi_2\psi_2}g_{\phi_2\phi_2}}
(D_2\nu_2 g_{\phi_2\phi_2}- C_2
\omega_2 g_{\psi_2\psi_2})]
\end{eqnarray}
This differential equation determines the most general evolutions of
$\theta$'s. As it is clear from the above, solving for general
$\theta$ is quite hard. Hence, in what follows we will analyze it in
some special situations.
\subsection{Magnon solutions in $R \times S^{3}_1$}
Let us start with the situation when $\theta'_2=h'_2=g'_2=0 \  ,
\omega_2=\nu_2=0$. For our convenience, we set $D_1 = \lambda_1 d$,
$C_1 = \lambda_1 c$ and $\kappa^2-v_r^2-v_y^2 = \lambda_1 \gamma$.
Then, the above equation takes the form
\begin{eqnarray}
\theta_1 '^2 &=& \frac{1}{(\alpha^2-\beta^2)^2}
\left[(\alpha^2+\beta^2) \gamma -\alpha^2 (\nu_1^2 - \omega_1^2) - 2 \alpha (d \nu_1 + c\omega_1)
\right. \nonumber \\
&& \qquad \left. - \frac{(\alpha \omega_1-c)^2}{\sin^2 \theta_1}
- \frac{d^2}{\cos^2\theta_1}\right] \ .
\end{eqnarray}
Let us write the above differential equation in the following form
\begin{equation}
\theta_1'^2= A^2-\frac{B^2}{\sin^2\theta_1}
-\frac{{d'}^2}{\cos^2\theta_1}  \ ,
\end{equation}
where \bea A^2 &=& \frac{1}{(\alpha^2 -
\beta^2)^2}\left[(\alpha^2+\beta^2)
\gamma -\alpha^2 (\nu_1^2 - \omega_1^2)
- 2 \alpha (d \nu_1 +
c\omega_1)\right], \nn B^2 &=&
\frac{1}{(\alpha^2 -
\beta^2)^2}\left[(\alpha
\omega_1-c)^2\right] \ ,  {d'}^2 =
\frac{d^2}{(\alpha^2 -\beta^2)^2} \ .
\eea
 Then  \be \theta_1' = A \
\frac{\sqrt{(\sin^2\theta_1-\sin^2\theta_{min})
(\sin^2\theta_{max}-\sin^2\theta_1)}}
{\sin\theta_1 \cos \theta_1} \  , \ee
where \bea \sin^2\theta_{max} &=&
\frac{(A^2+B^2-{d'}^2)+\sqrt{(A^2+B^2-{d'}^2)^2-4A^2B^2}}
{2A^2} \ , \nonumber \\
\sin^2\theta_{min} &=&
\frac{(A^2+B^2-{d'}^2)-\sqrt{(A^2+B^2-{d'}^2)^2-4A^2B^2}}
{2A^2} \   . \eea
Let us try to find
the solution when
$\sin^2\theta_{max}=1$. This occurs
when
\begin{eqnarray}
D_1=0 \ , \quad {\rm with} \
\sin\theta_{min}=\frac{B}{A} \ .
\end{eqnarray}
The final form of $\th_1 '$ is
\begin{equation}
\theta_1 '=A \frac{\sqrt{\sin^2\theta_1-
\sin^2\theta_{min}}}{\sin\theta_1} \ .
\end{equation}
Note that the range of $\th_{1,min} \le
\th_1 \le \frac{\pi}{2}$ corresponds to
the half of a magnon-like solution. For
the convenience, we call one element of
the solitonic string solution having
the magnon shape as a magnon-like
solution. Actually, the solution
obtained here is a combination of these
magnon-like solution. So the number of
the magnon-like solution contained in
the solitonic string solution is given
by the followings. If the range of the
string world sheet is given by
$-\frac{\pi}{2} \le \s \le
\frac{\pi}{2}$ for an open string, the
number of magnon-like solution is
determined by \bea \la{num1} \pi =
\int_{-\pi/2}^{\pi/2} d \s = \frac{2
n}{\a} \int_{\th_{1,min}}^{\pi/2}
\frac{d \th_1}{\th_1 '}, \eea where $n$
means the number of the magnon-like
solutions. After calculating the last
equation, the number of the magnon-like
solutions is given by $n= \a A$.

From now on, we will concentrate on the
conserved quantities for magnon-like
solutions which will give a dispersion
relation for one magnon-like solution.
Using the definitions of the conserved
charges (\ref{CGcon}) they can be
rewritten as the integral form over
$\theta_1 $ \bea
P_t &=& \frac{1}{\pi \a'} \frac{\k}{\a} \ I ,\\
P_y &=& \frac{1}{\pi \a'} \frac{v_y}{\a} \ I ,\\
P_r &=& \frac{1}{\pi \a'} \frac{v_r}{\a} \ I ,\\
P_{\ph_1} &=& - \frac{\l_1}{\pi \a'} \frac{1}{\a^2-\b^2}
\left( \frac{\b}{\a} c + \a \n_1 \right) \ I ,\\
P_{\ps_1} &=& - \frac{\l_1}{\pi \a'} \frac{c - \a \o_1}{\a^2-\b^2}
\left( I - I' \right) ,
\eea
where
\bea
I &=& \int_{\th_{min}}^{\pi /2} d \th_1  \
 \frac{1}{\th_1 '} = \frac{\pi}{2A} \  , \nn
I' &=& \int_{\th_{min}}^{\pi /2} d \th_1  \
 \frac{1}{\th_1 ' \sin^2 \th_1 }  = \frac{\pi}{2B} \ .
\eea The angle difference in the
$\ph_1$-direction is given by \be \D
\ph_1 = \frac{2}{\a^2-\b^2} \left[
(c-\a \o_1) I' + (\a \o_1 + \b \n_1) I
  \right] \ .  \ee
To obtain the dispersion relation, we
first consider the following quantity
\bea P_t^2 - P_r^2 -P_y^2 &=&
\frac{1}{(\pi \a' \a)^2} (\k^2 -v_r^2
-v_y^2) \ I^2 \nn &=& - \frac{1}{(\pi
\a' \a)^2} \frac{\l_1 \n_1 c}{\b} \
I^2 \, \eea where the Virasoro
constraints are used in the last
equation. From the definitions of
charges, $\n_1 c \ I^2$ in the above
equation can be determined in terms of
other charges and the angle difference
\be -  \frac{\l_1 \n_1 c}{\b} \  I^2 =
\frac{(\pi \a' \a)^2}{\l_1} \left[
P_{\ph_1}^2 + \frac{\b^4 -\a^4}{2 \a^2
\b^2} P_{\ph_1}^2 - \frac{2 \b}{\a}
P_{\ph_1} (P_{\psi_1} - T_1 \D \ph_1) +
(P_{\psi_1} - T_1 \D \ph_1)^2 \right] .
\ee Finally, we obtain the dispersion
relation \bea P_t^2 - P_r^2 -P_y^2 &=&
\frac{1}{\l_1} \left[ P_{\ph_1}^2 +
\frac{\b^4 -\a^4}{2 \a^2 \b^2}
P_{\ph_1}^2 - \frac{2 \b}{\a} P_{\ph_1}
(P_{\ps_1} - T_1 \D \ph_1) + (P_{\ps_1}
- T_1 \D \ph_1)^2 \right] \nn &=&
\frac{1}{\l_1} \left[
\left(1-\frac{\a^4+\b^4}{2\a^2 \b^2}
\right) P_{\ph_1}^2 + (\frac{\b}{\a}
P_{\ph_1} + T_1  p  -  P_{\ps_1} )^2
\right] , \eea where $T_1 =
\frac{\l_1}{2 \pi \a'}$ and we have
identified the angle difference $\D
\ph_1$ with the world sheet momentum
$p$.

\subsection{Magnon solutions on $R\times S^{3}_1 \times S^{3}_2$}

The equations of motion for $\ph_i$ and
$\ps_i$ ($i = 1,2$) are summarized as
\bea h_i' &=& \frac{1} {\lambda_i
(\alpha^2-\beta^2) \sin^2\theta_i}
(C_i- \alpha \lambda_i
\omega_i\cos^2\theta_i+ \beta \lambda_i
\nu_i \sin^2\theta_i) \ , \nn g_i '&=&
\frac{1}{ \lambda_i (\alpha^2-\beta^2)
\cos^2\theta_i} (D_i + \alpha \lambda_i
\nu_i  \cos^2\theta_i + \beta \lambda_i
\omega_i \cos^2\theta_i ) \ . \eea
($\ref{mas1}$) becomes \be
\sum_{i=1}^{2} \l_i \th_i '^{2} =
\frac{\alpha^2+\beta^2}{(\alpha^2-\beta^2)^2}
(\kappa^2-v_r^2-v_y^2) - \sum_{i=1}^{2}
K_i (\th_i), \ee where \bea K_i (\th_i)
&=&  \frac{1}{(\alpha^2-\beta^2)^2}
\left[  \frac{\lambda_i
\alpha^2}{\sin^2\theta_i}( \nu_i^2
\sin^2\theta_i + \omega_i^2
\cos^2\theta_i) +
[\frac{C_i^2}{\lambda_i \sin^2\theta_i}
+\frac{D_i^2}{\lambda_i  \cos^2\theta_i} \right.  \nonumber \\
&& \qquad \left. + \frac{2 \alpha}{\sin^2\theta_i}
( D_i\nu_i \sin^2 \theta_i -
C_i\omega_i \cos^2\theta_i ) ] \right]  .
\eea
Without loss of generality, we can set
\be
 \l_i \th_i '^{2} +  K_i (\th_i) \equiv \l_i \G_i  \ ,
\ee
where $\G_i$ are some constants satisfying $\l_1 \G_1 + \l_2 \G_2 =
\frac{\alpha^2+\beta^2}{(\alpha^2-\beta^2)^2}
(\kappa^2-v_r^2-v_y^2)$.

When we set $D_i = 0$ and $C_i = \l_i c_i$, $\th_i'$ are given by
\be
\theta_i' = A_i \
\frac{\sqrt{(\sin^2\theta_i-\sin^2 \theta_{i,min})}}
{\sin\theta_i } ,
\ee
where
\bea
A_i^2 &=& \G_i - \frac{\alpha^2}{(\alpha^2-\beta^2)^2}
(\nu_i^2-\omega_i^2)+\frac{2\alpha c_i \omega_i}
{(\alpha^2-\beta^2)^2}
 \ , \nn
\sin^2 \theta_{i,min} &=& \frac{(\alpha \omega_i-c_i)^2}{A_i^2} .
\eea
In this background, the number of the magnon-like solution in each sphere is given
by the similar relation in \eq{num1}
\be
\pi = \frac{2 n_i}{\a} \int_{\th_{i,min}}^{\pi/2} \frac{d \th_i}{\th_i '} =
\frac{n_i \pi}{\a A_i},
\ee
where $n_i$ means the number of the magnon-like solution in $i$-th sphere. From the
above equation, the number of the magnon becomes $n_i = \a A_i$. If we set the ratio between
magnon-like numbers as $r \equiv \frac{n_2}{n_1}$, then this ratio is given by
\be
r = \frac{A_2}{A_1} \ .
\ee
From now on, we set a magnon-like solution as one in the first sphere, which corresponds
to $r$ magnon-like solutions in the second sphere.

The conserved charges for a magnon-like solution are
\bea
P_t &=& \frac{1}{\pi \a'} \frac{\k}{\a} \ I_1 ,\\
P_y &=& \frac{1}{\pi \a'} \frac{v_y}{\a} \ I_1 ,\\
P_r &=& \frac{1}{\pi \a'} \frac{v_r}{\a} \ I_1 ,\\
P_{\ph_i} &=& - \frac{\l_i}{\pi \a'} \frac{1}{\a^2-\b^2}
\left( \frac{\b}{\a} c_i + \a \n_i \right) \ I_i ,\\
P_{\ps_i} &=& - \frac{\l_i}{\pi \a'} \frac{c_i - \a \o_i}{\a^2-\b^2}
\left( I_i - I_i' \right) \ , \\
\D \ph_i &=& \frac{2}{\a^2-\b^2} \left[ (c_i -\a \o_i) I_i' + (\a
\o_i + \b \n_i) I_i \right] , \eea where \bea I_1 &=&
\int_{\th_{1,min}}^{\pi /2} d \th_1  \  \frac{1}{\th_1 '} =
\frac{\pi}{2A_1} , \nn I_1' &=& \int_{\th_{1,min}}^{\pi /2} d \th_1
\  \frac{1}{\th_1 ' \sin^2 \th_1 } = \frac{\pi}{2B_1}\ , \nn I_2 &=&
r \int_{\th_{2,min}}^{\pi /2} d \th_2  \  \frac{1}{\th_2 '} =
\frac{r \pi}{2A_2}
 , \nn
I_2' &=& r \int_{\th_{2,min}}^{\pi /2}
d \th_2  \  \frac{1}{\th_2 ' \sin^2
\th_2 } = \frac{r \pi}{2B_2}\ , \eea
with $B_i^2 = (\alpha \omega_i-c_i)^2$.
Then the relation (\ref{const})
 $\kappa^2 -
v_r^2 - v_y^2 =\frac{1}{\b} (-\nu_1 C_1
-\nu_2 C_2)$
 can be rewritten in terms
of  charges \bea P_t^2 - P_r^2 -P_y^2
&=& \sum_{i=1}^2 \frac{1}{\l_i} \left[
P_{\ph_i}^2 + \frac{\b^4 -\a^4}{2 \a^2
\b^2} P_{\ph_i}^2 - \frac{2 \b}{\a}
P_{\ph_i} (P_{\ps_i} - T_i \D \ph_i)
\right. \nn && \left. \qquad \qquad +
(P_{\ps_i} - T_i \D \ph_i)^2 \right] \nn
&=& \sum_{i=1}^2 \frac{1}{\l_i} \left[
\left(1-\frac{\a^4+\b^4}{2\a^2 \b^2}
\right) P_{\ph_i}^2 + \left(\frac{\b}{\a}
P_{\ph_i} + T_i  \D \phi_i  -  P_{\ps_i} \right)^2
\right] , \eea where $T_i =
\frac{\lambda_i}{2\pi\alpha'}$ with
$i=1,2$. This corresponds to the
dispersion relation for the string
soliton on the I-brane background when the it moves simultaneously 
on both the  spheres, and this does not depend on the previous
parameterization, $\G_i$.

\section{Discussion} In this paper we have studied the solutions for
rotating strings in the background
generated by a 1+1 dimensional
intersection of two stacks of five
branes in type IIB string theory. We
have solved the motion of rotating
string in this background and have
analyzed the dispersion relation among
various conserved charges. We have
taken advantage of the fact there
exists a parameter space where the
motion in the two spheres effectively
decoupled, and one could study the
single magnon like solution in this
background. Knowing the results of the
present paper, it would be tempting to
study the corresponding states in the
dual theory exactly in case of the
AdS$_5\times$ S$^5$ background. It
would certainly be interesting to check
whether these magnon solutions are BPS
from the bulk theory view point, which
will give us clue about the boundary
operators. We wish to come back to this
issue in future.

\vskip .2in \noindent {\bf Acknowledgements:} KLP would like to
thank the hospitality at Institute of Physics, Bhubaneswar, India
where a part of this work was done. This work was partially
supported by the Science Research Center Program of the Korea
Science and Engineering Foundation through the Center for Quantum
Spacetime (CQUeST) of Sogang University with grant number R11 - 2005
- 021. The work of JK  was supported  by the Czech Ministry of
Education under Contract No. MSM 0021622409.


\newcommand{\np}[3]{Nucl. Phys. {\bf B#1}, #2 (#3)}
\newcommand{\pprd}[3]{Phys. Rev. {\bf D#1}, #2 (#3)}
\newcommand{\jjhep}[3]{J. High Energy Phys. {\bf #1}, #2 (#3)}

\end{document}